\documentclass[10pt]{iopart}

\usepackage{cite}

\expandafter\let\csname equation*\endcsname\relax
\expandafter\let\csname endequation*\endcsname\relax
\usepackage{amsmath, amsbsy}
\usepackage[T1]{fontenc}
\usepackage{xurl}
\usepackage[unicode=true, bookmarks=true, bookmarksnumbered=true, bookmarksopen=true, bookmarksopenlevel=1, breaklinks=false,pdfborder={0 0 0}, backref=false, colorlinks=true]{hyperref}
\hypersetup{citecolor=blue,linkcolor=blue,urlcolor=blue}

\usepackage[dvipsnames]{xcolor} 
\usepackage{lipsum,multicol} 

\usepackage [english]{babel}
\usepackage [autostyle, english = american]{csquotes}
\MakeOuterQuote{"}

\usepackage{graphicx}

\begin{document}

\title{Quantifying resonant drive in resistive perturbed tokamak equilibria}

\author{M. Pharr$^1$,
N.C. Logan$^{1,2}$,
C. Paz-Soldan$^{1,2}$,
J.-K. Park$^{3}$
}
\address{
$^1$ Department of Applied Physics and Applied Mathematics, Columbia University, New York, New York 10027, USA

$^2$ Columbia Fusion Research Center, Columbia University, New York, New York 10027, USA

$^3$ Department of Nuclear Engineering, Seoul National University, Seoul 08826, Republic of Korea
 }
\ead{mcp2198@columbia.edu}

\vspace{10pt}
\begin{indented}
\item[]\today
\end{indented}

\begin{abstract}

Resonant drive in tokamaks is routinely quantified using a variety of different metrics that target different aspects of a resonant response to an external perturbation. Two of the most direct metrics, $\Delta_{mn}$ and $b_{pen}$, are widely used but their relative behavior was previously uncharacterized. This work examines how these metrics representing the shielding current and penetrated field relate in resistive perturbed tokamak equilibria using asymptotically matched solutions with a resistive MHD inner layer model in GPEC. $b_{pen}$ scales with Lundquist number as $S^{-2/3}$ until saturation at low $S$, and $\Delta_{mn}$ remains consistent with its ideal definition but is affected by global kink structure. Both metrics are shown to yield closely similar dominant coupling modes within the same resistive model. However, the resistive physics shifts this dominant mode spectrum to lower poloidal mode numbers $m$ in a low-rotation ITER equilibrium. This alteration is predicted to be observable in experiment in the form of optimal relative phasings of resonant magnetic perturbation coils.

\end{abstract}

%
%
\submitto{\NF}
%
\maketitle
%
\ioptwocol

\section{\label{sec:intro} Introduction}

\begin{figure}
\includegraphics[width=\linewidth]{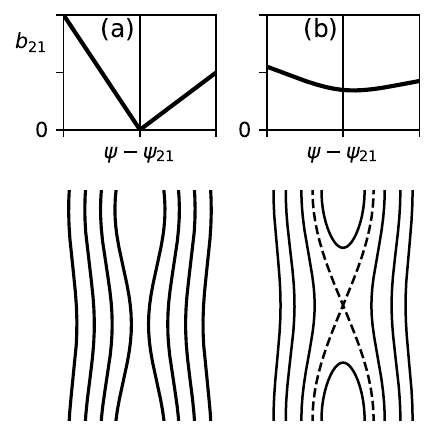}\protect\caption{A schematic showing the preservation of magnetic field topology in an ideal plasma (a) compared to the opening of an island as magnetic field lines reconnect (b) in the presence of a 2/1 radial field perturbation. Vertical direction in the lower island schematics is increasing poloidal coordinate and horizontal direction is radial, the same as the upper plots. \label{fig:island_opening}}
\end{figure}

Resonant magnetic perturbations (RMPs) are 3D fields in tokamaks that can arise due to error fields from coil misalignments or ferritic materials as well as intentional perturbations for the correction of error fields or suppression of edge-localized modes (ELMs). These fields affect otherwise axisymmetric tokamaks in a variety of ways, including driving magnetohydrodynamic (MHD) modes \cite{kargerInfluenceResonantHelical1974,pironErrorFieldDetection2024}, braking plasma rotation through neoclassical toroidal viscosity (NTV) \cite{shaingSymmetryBreakingInducedTransport2001,shaingApproximateAnalyticExpression2010,callenEffects3DMagnetic2011}, increasing transport \cite{nazikianPedestalCollapseResonant2021,callenResonantMagneticPerturbation2012}, and suppressing edge-localized modes \cite{zohmEdgeLocalizedModes1996,paz-soldanPlasmaPerformanceOperational2024,evansSuppressionLargeEdgeLocalized2004}. Fig.~\ref{fig:island_opening} sketches the effects of a resonant magnetic perturbation at a rational surface, showing the differences in magnetic field topology enforced by the presence or absence of resistivity. These effects are modeled using a variety of approaches, including linear ideal and non-ideal MHD force-free states solvers such as the GPEC suite \cite{parkComputationThreedimensionalTokamak2007,parkSelfconsistentPerturbedEquilibrium2017,wangModelingResistivePlasma2020}, linear MHD eigenvalue solvers such as those in the MARS family \cite{liuFeedbackStabilizationNonaxisymmetric2000,liuToroidalSelfconsistentModeling2008}, as well as high-performance nonlinear extended MHD codes such as NIMROD \cite{sovinecNonlinearMagnetohydrodynamicsSimulation2004}, JOREK \cite{huysmansMHDStabilityXpoint2007,czarnyBezierSurfacesFinite2008,hoelzlJOREKNonlinearExtended2021}, and M3D-C1 \cite{ferraroM3DC12018,jardinM3DC1ApproachSimulating2008,ferraroCalculationsTwofluidMagnetohydrodynamic2009,breslauPropertiesM3DC1Form2009}. Modeling efforts using these codes have seen many successful applications, including successful prediction of 3D coil currents that induce ELM suppression \cite{park3DFieldPhasespace2018,parkOptimizing3DMagnetic2023}, accurate models of optimal single- and multiple-set coil currents for the elimination of core-resonant error fields that can cause mode locking \cite{paz-soldanSpectralBasisOptimal2014,paz-soldanImportanceMatchedPoloidal2014,parkErrorFieldCorrection2011}, and quasi-symmetric kinetic optimization of perturbed fields that reduce NTV braking \cite{parkQuasisymmetricOptimizationNonaxisymmetry2021}, among others. Every model-forward study of resonant 3D fields uses quantitative metrics of resonant mode drive, which we refer to as "resonant metrics."

There are many resonant metrics that serve varying purposes in the community. The X-point displacement is one such quantity, which was adopted as a metric when studies \cite{liuModellingPlasmaResponse2011,liuComparativeInvestigationELM2017} showed that the maximum displacement of the outer surface of a modeled plasma correlates with ELM suppression in several tokamaks. This metric is naturally used because it can both be retrieved from spectral simulations \cite{weisbergOptimizingMultimodalNonaxisymmetric2019} and be compared to experimental data, but it has been shown to be sensitive to the equilibrium reconstruction \cite{loganMetricsExtrapolationResonant2025}. One old metric used in design work for ITER is the 3-mode metric \cite{scovilleMultimodeErrorField2003,henderChapter3MHD2007,amoskovFourierAnalysis3D2004}, which arbitrarily weights the first three ($m=1-3$) Fourier harmonics of the resonant vacuum field in a geometric average. This has the advantage of referring to a resonant quantity that directly drives penetration, but the omission of the plasma response was a critical flaw, as well as its dependence on the coordinate system \cite{parkSpectralAsymmetryDue2008,schafferStudyInvesselNonaxisymmetric2008} and its demonstrated experimental inadequacy \cite{paz-soldanImportanceMatchedPoloidal2014,paz-soldanSpectralBasisOptimal2014}. The island overlap width (IOW) and vacuum island overlap width (VIOW) are other commonly used metrics --- however, since they have a discontinuous relationship with perturbation amplitude, they are usually seen alongside their respective Chirikov parameters \cite{fenstermacherEffectIslandOverlap2008,kirkMagneticPerturbationExperiments2011}. The IOW requires a plasma response model, which can either come from a resistive code which gives a finite penetrated resonant field, or from an ideal code in which the penetrated resonant field is identically zero and the "effective" island width is calculated from the field derivative discontinuity $\Delta_{mn}$ \cite{boozerPerturbedPlasmaEquilibria2006}. The response-inclusive Chirikov parameter too has drawbacks, as the shielded flux is not equivalent to the penetrated flux which actually correlates with island width in resistive models. 

$\Delta_{mn}$ itself is another commonly used resonant metric, and is now (in the form of the associated shielded resonant flux) used in ideal analyses of error field correction (EFC) in DIII-D \cite{paz-soldanSpectralBasisOptimal2014,paz-soldanImportanceMatchedPoloidal2014}, ITER \cite{parkErrorFieldCorrection2008,parkMDC19ReportAssessment2017,amoskovAdvancedComputationalModel2018,bandyopadhyayMHDDisruptionsControl2025,mcintoshITERsAssemblyTolerances2023,pharrErrorFieldPredictability2024,baiTimeVariationError2025}, and SPARC \cite{loganSPARCTokamakError2026}, among others. 
$\Delta_{mn}$ has the advantage of being proportional to the shielding current at a rational surface, and therefore essentially a direct coefficient for the magnitude of the resonant response. 
However, $\Delta_{mn}$ is only well-defined in ideal MHD, where there is a discontinuity in the magnetic field at rational surfaces. Resistive and other non-ideal MHD formulations allow for a finite penetrated field at rational surfaces, which smooths this discontinuity. 
Several quantities derived from $\Delta_{mn}$ have also been used. One such is the average magnitude of shielded flux in the plasma edge, which was used to accurately predict the multi-dimensional phase space of ELMy, ELM-suppressed, and mode-locked states in KSTAR \cite{park3DFieldPhasespace2018}. More robust than this simple average is the dimensionless, singular value decomposed (SVD) inner product quantity $\delta$ known as the dominant mode overlap \cite{parkErrorFieldCorrection2011,loganMetricsExtrapolationResonant2025}. The overlap $\delta$ has seen applications in identifying optimal EFC currents \cite{paz-soldanImportanceMatchedPoloidal2014} and scaling-law-based \cite{loganRobustnessTokamakError2020,loganEmpiricalScaling22020} design evaluations \cite{pharrErrorFieldPredictability2024,loganSPARCTokamakError2026}. 
Another direct metric is the penetrated resonant field $b_{pen}^{res}$ itself. Specifically, this is the plasma response inclusive pitch-resonant surface-normal field, $\frac{1}{\left|\nabla \psi\right|}\vec b_{mn} \cdot \nabla \psi$. $b_{pen}^{res}$ can be identified in resistive codes like M3D-C1 and MARS-F, and its use has been reported in evaluations of resonant drive in EAST \cite{yangToroidalModeling12018,wangDensityScaling12018,yangModellingPlasmaResponse2016}, DIII-D \cite{liuComparativeInvestigationELM2017,weisbergPassiveDeconfinementRunaway2021,lyonsEffectRotationZerocrossing2017}, and ASDEX Upgrade \cite{liuELMControlRMP2016,liuModellingPlasmaResponse2011}. Having a perturbed pitch-resonant surface-normal field at a rational surface is synonymous with having an open island; therefore, its magnitude is a very direct measure for resonant drive. This is a natural improvement over using the vacuum perturbed pitch-resonant surface-normal field.

\begin{figure}
\includegraphics[width=\linewidth]{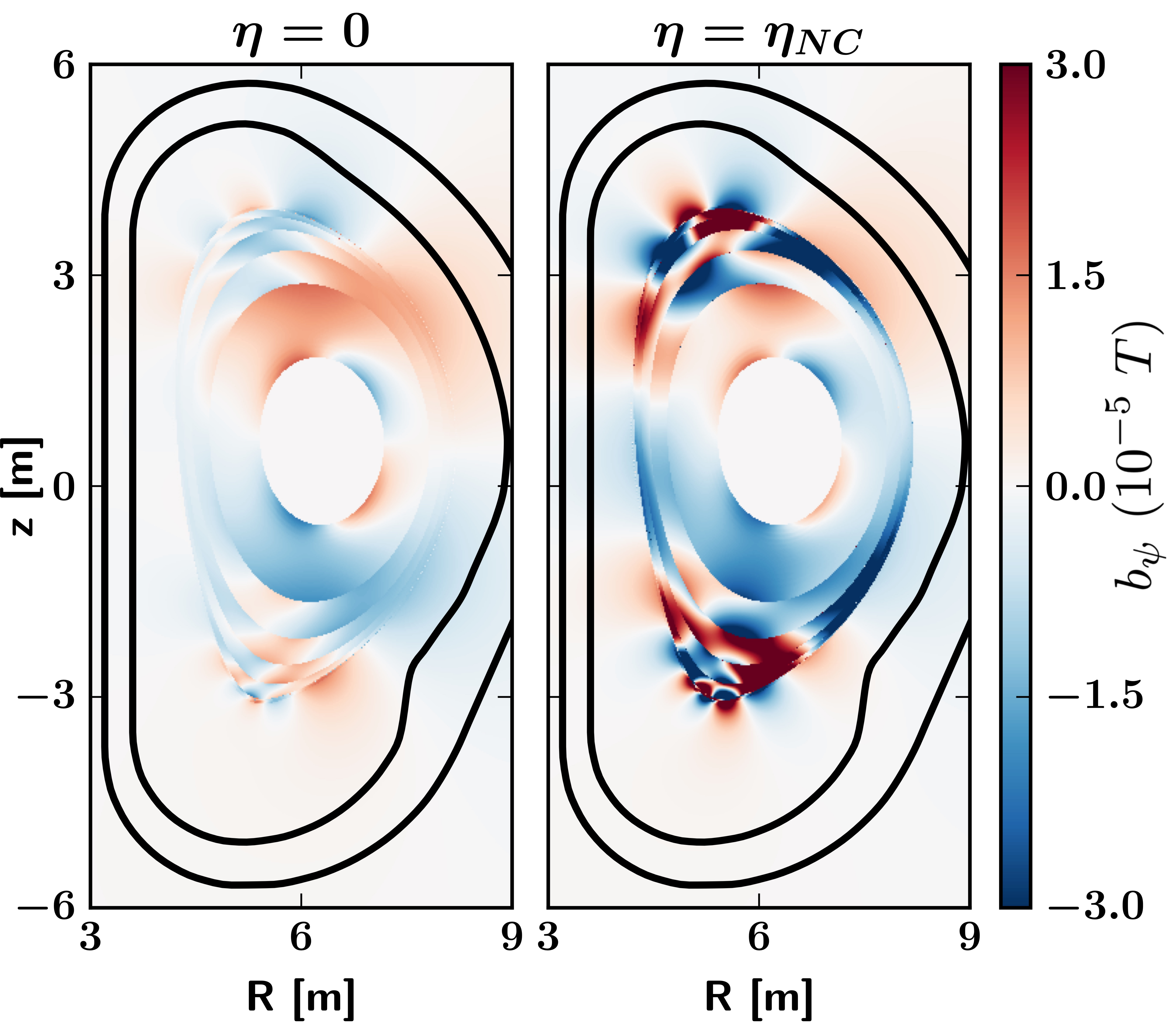}\protect\caption{Ideal (left) and resistive (right) total perturbed fields in ITER when applying a 1 kA perturbation via the middle EFC coil, including the plasma response. The resistive plasma response, using a neoclassical resistivity $\eta_{NC}$ from Jardin \cite{jardinTSCSimulationOhmic1993}, is stronger and has noticeably different structure compared to the ideal response. \label{fig:models}}
\end{figure}

The sheer number of metrics in use makes it difficult to compare experimental analyses and modeling insights between studies that use substantively different metrics. Many of the aforementioned metrics have been studied in the context of RMP ELM suppression \cite{loganMetricsExtrapolationResonant2025}, but beyond the range of how directly different metrics quantify drive, the lack of understanding of the relationship between the shielded resonant field $b_{sh}^{res}$ (derived from $\Delta_{mn}$) and the resonant penetrated field $b_{pen}^{res}$, the two most direct metrics, is a missed opportunity to understand the validity of ideal models of resonant behavior. In this paper, we establish an understanding of how resonant metrics based on shielding current and penetrated field relate to each other in ideal and resistive perturbed equilibria calculated using the resistive GPEC model \cite{wangModelingResistivePlasma2020}. Resistivity can significantly alter plasma response to 3D fields, as seen in Fig.~\ref{fig:models}, and thus it is important to be able to determine when this alteration significantly changes the tearing drive.

This paper takes the following structure: first, in Sec.~\ref{sec:model}, we describe the resistive GPEC model and the physical assumptions made. Sec.~\ref{sec:Dmn} recalls the definition of the resonant field derivative discontinuity $\Delta_{mn}$ in ideal MHD and discusses how to choose a finite jump width for calculating $\Delta_{mn}$ in resistive models. In Sec.~\ref{sec:behavior}, we examine how the shielded field $b_{sh}^{res}$ and penetrated field $b_{pen}^{res}$ behave across several equilibria as resistivity is varied. Sec.~\ref{sec:efc} examines the impact of resistive physics on dominant mode spectra, showing that while the two metrics yield closely similar dominant mode spectra when evaluated within the same resistive model, the inclusion of resistive physics can modify the dominant mode coupling spectrum relative to ideal calculations in some regimes. We also demonstrate experimentally observable modifications to EFC coil phasings in ITER. Finally, we summarize the implications of these results for the use of resonant metrics in resistive perturbed equilibria and discuss future work in Sec.~\ref{sec:summary}.

\section{\label{sec:model} The Resistive GPEC Model}

The recently open-sourced GPEC framework \cite{parkComputationThreedimensionalTokamak2007,parkGeneralizedPerturbedEquilibrium2018} uses force-free solutions of the $\delta W$ Euler-Lagrange equation, also known as the toroidal Newcomb equation, from the stability code DCON as a basis to build perturbed equilibria. Different variations of the DCON code exist that employ different physics. In this work, we will use an ideal Euler-Lagrange equation calculated via Resistive DCON's (RDCON's) Galerkin method\cite{glasserComputationResistiveInstabilities2016a,wangModelingResistivePlasma2020}. 

The Galerkin method splits the integration up into an array of finite elements, each employing a number of Hermite polynomial basis functions through which the solution is represented. Basis functions that are singular at the rational surfaces are needed in the neighborhood of the rational surfaces, and are thus included in cells adjacent to the rational surface, known as resonant cells. There are also several different kinds of other cells with additional basis functions, chosen to achieve $\mathcal{C}^1$ continuity in solution functions.

This leads to the direct calculation of asymptotic coefficients in the limit $| \psi-\psi_r | \rightarrow 0$, as these are just the coefficients of the largest resonant basis function. 
Then, the RMATCH module \cite{wangModelingResistivePlasma2020} is employed to solve the Glasser-Green-Johnson (GGJ) inner layer model \cite{glasserResistiveInstabilitiesGeneral1975,glasserResistiveInstabilitiesTokamak1976,glasserNumericalSolutionResistive1984} and calculate the asymptotic coefficients in the opposing limit $| \psi-\psi_r | \rightarrow \infty$. RMATCH constructs resistive solutions by matching these coefficients between the ideal outer and resistive inner layers. This is an acceptable approximation of the methods of straight-through codes like MARS-F, which employ resistive physics throughout the plasma, as resistive terms in the MHD equations are small save for a neighborhood about rational surfaces. This is qualified by the fact that asymptotic matching approaches base themselves on the assumption that the inner layer is small, so this approximation fails in plasmas with very low Lundquist numbers $S$. However, as the community further approaches the achievement of steady-state confinement of burning plasmas, this approximation only becomes more valid as resistivities fall with the lower collisionality of a high-temperature plasma.

\begin{figure}[ht]
    \includegraphics[width=\linewidth]{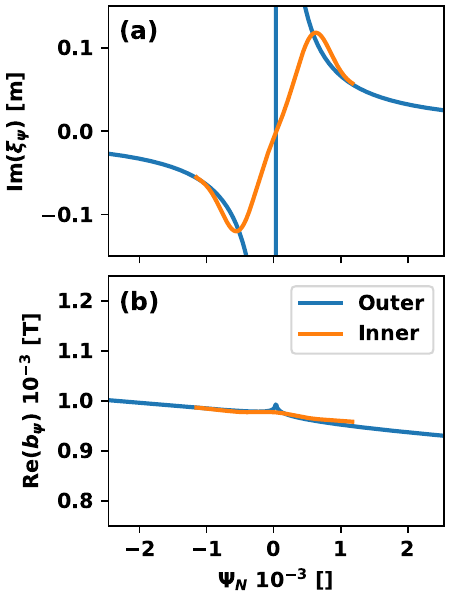}
    \caption{Inner and outer layer structures shown from LAR case (a) in Table~\ref{tab:equilibria} at the $q=2$ surface, (a) $\xi^{m=2}_\psi$ and (b) $b^{m=2}_\psi$.}\label{fig:innerouter}
\end{figure}

While the relatively old GGJ model is superseded by much more extensive inner layer models today \cite{parkParametricDependenciesResonant2022,fitzpatrickInvestigationNeoclassicalTearing2025,dudkovskaiaDriftKineticTheory2021}, this inner layer model can be switched out with other models for a more sophisticated representation of the global mode. The GGJ model solves the perturbed resistive MHD equations in the inner layer:

\begin{subequations}\label{eq:ggj}
\begin{align}
\gamma^2 \rho\,\vec{\xi} &= -\nabla p + \vec{j}\times \vec{B} + \vec{J}\times \vec{b} \label{eq:ggj_a}\\
\gamma\,\vec{b} &= \gamma \nabla\times\!\left( \vec{\xi}\times \vec{B} \right) - \nabla\times\!\left( \eta\,\vec{j} \right) \label{eq:ggj_b}\\
p &= -\vec{\xi}\cdot \nabla P - \Gamma P\,\nabla\cdot \vec{\xi} \label{eq:ggj_c}\\
\mu_{0}\,\vec{j} &= \nabla\times \vec{b}. \label{eq:ggj_d}
\end{align}
\end{subequations}

\begin{figure}
    \includegraphics[width=\linewidth]{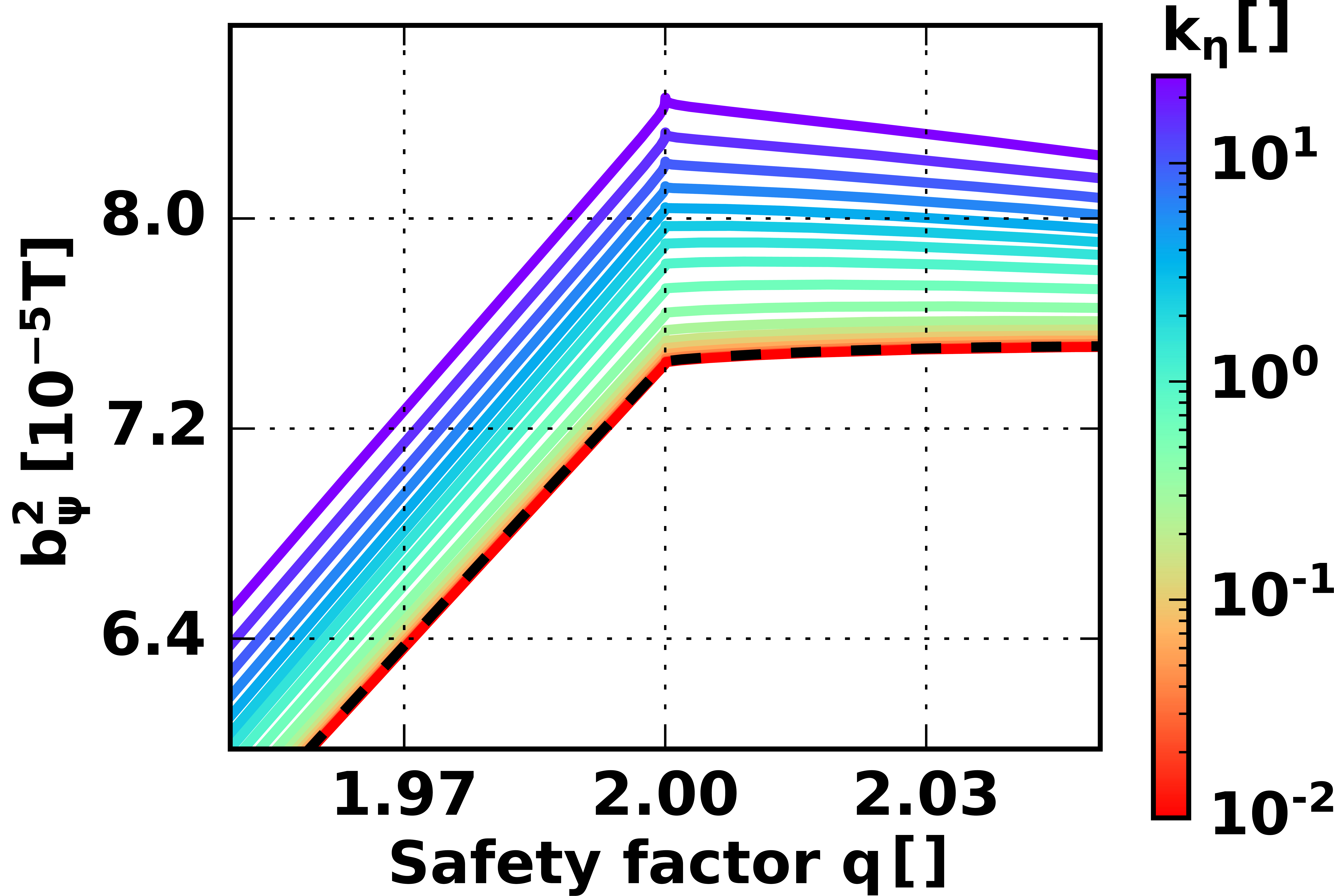} \caption{The structure of resistive GPEC solutions varies with $\eta$ --- for small perturbations from ideal MHD, $b_{pen}^{res} \sim S^{-2/3} \sim \eta^{2/3}$. The ideal solution is shown as a black dashed line. $k_\eta$ is a constant multiplier that scales resistivity on all surfaces, with $k_\eta=1$ signifying neoclassical resistivity. Solutions shown are in a DIII-D-like equilibrium, detailed in Table~\ref{tab:equilibria} and Fig.~\ref{fig:equilibria}.}\label{fig:res_scans_b}
\end{figure}

As resistivity diffusively consumes flux over time, the concept of a perturbed equilibrium is slightly different from the ideal case; we need a growth rate $\gamma$ to have non-singular solutions of Eq.~\ref{eq:ggj}. Thus, for a perturbed equilibrium, $\gamma$ is taken to be the Doppler shifted frequency, $\gamma = i n \Omega = 2\pi i n f$. This model is similar to that of the MARS-F code \cite{liuFeedbackStabilizationNonaxisymmetric2000}. The primary difference is that in our model, resistivity is only present where its magnitude is significant in solving for a perturbed resistive equilibrium, that being near the rational surfaces where otherwise large terms limit to zero. Details of the numerical methods of resistive GPEC can be found in \cite{glasserComputationResistiveInstabilities2016a,glasserAsymptoticSolutionsConvergence2020,wangModelingResistivePlasma2020}. Furthermore, we note that all solutions for magnetic field presented in this paper will be the outer region solutions unless otherwise specified, as resistive magnetic field solutions represent just a smoothing of magnetic field over the rational surface as sketched in Fig.~\ref{fig:island_opening}, which is already present in the ideal solutions due to the coefficient matching process, as shown by Fig.~\ref{fig:innerouter}. The singularity is seen to be smoothed over in Fig.~\ref{fig:innerouter}(a), but the magnetic field derived from this Lagrangian displacement solution is already damped by a factor of $(m-nq)$ and is therefore already smoothed in Fig.~\ref{fig:innerouter}(b). Final outer region solutions after matching display basic properties of resistive physics, including convergence to ideal solutions as resistivity goes to zero, and increasing penetrated resonant field with increasing resistivity, as shown in Fig.~\ref{fig:res_scans_b}.

Resistive GPEC can be sensitive to equilibrium truncation, as well as to inputs like number of Galerkin elements. It is especially sensitive for large values of rotation $\Omega$. Numerical outputs, particularly of the inner layer solution, must be checked carefully to ensure a valid solution was reached. Asymptotic matching relies on the assumption that the inner layer width is small compared to the outer layer, and does not determine tearing mode stability as $\gamma = i n \Omega$ is assumed. Our main focus in this work is on results obtained from the outer region solutions computed via matching to the inner layer.

\section{\label{sec:Dmn} Field derivative discontinuity and choosing a discrete jump width} 

In ideal MHD, magnetic field topology is immutable and field lines are prevented from reconnecting due to the absence of resistive diffusivity. Thus, in order to prevent an island from opening on a rational surface, a shielding current appears in the presence of linear perturbations that contributes an equal and opposite field on the rational surface, identically fixing the resonant penetrated field at zero. This current was shown by Boozer and Nührenberg \cite{boozerPerturbedPlasmaEquilibria2006} to be proportional to the quantity $\Delta_{mn}$:

\begin{subequations}\label{eq:delta_mn}
\begin{align}
j_{mn} &= -i \frac{J_{c}}{m}\Delta_{mn}\\
\Delta_{mn} &\equiv \left[ \frac{ \partial }{ \partial \psi } \frac{{\vec{b}\cdot \nabla \psi}}{\vec{B}\cdot \nabla \phi} \right]_{mn}
\end{align}
\end{subequations}

\noindent where $J_{c}$ is the amplitude of a characteristic current determined by equilibrium quantities and $\vec{B}\cdot \nabla \phi$ is a coordinate Jacobian, making the quantity dimensionless. $\Delta_{mn}$ is well-defined in ideal MHD and can be simply calculated in ideal codes by taking the difference in the radial field derivative across a small width whose magnitude is determined only by the numerical sensitivity of the solver near the singular point.  

In the presence of finite resistivity, however, magnetic field lines are able to reconnect, which diffuses the sharp current sheet over a finite width and allows a finite penetrated resonant field at the rational surface. This diffusion suppresses the MHD singularity at the rational surface and results in a continuous derivative $\frac{ \partial }{ \partial \psi } \vec{b}\cdot \nabla \psi$, as seen in Fig.~\ref{fig:island_opening}. Therefore, $\Delta_{mn}$ can no longer be interpreted as a discontinuity; this motivates the redefinition of $\Delta_{mn}$ as the integrated shielding current. 

Adopting Boozer's notation, we use a magnetic coordinate system $\left( \psi,\theta_{m},\phi \right)$ at a surface where $q(\psi_{mn}) = m/ n$ has field lines on each surface with the label $\alpha \equiv \theta_{m} - \phi/q$. What was a perturbed parallel surface current in ideal MHD thickens,

\begin{align}
\begin{split}
\vec{j}_{\parallel}=j_{mn}e^{-im\alpha}\delta(\psi&-\psi_{mn}) \vec{B} \rightarrow \\
&j_{mn}e^{-im\alpha}f(\psi-\psi_{mn}) \vec{B}
\end{split}
\end{align}

\noindent where the integrable distribution $f$ has the property $\int_{-\delta\psi/2}^{\delta\psi /2} f(\psi) /\left| \nabla \psi \right| \, d\psi = 1 /\left| \nabla\psi \right|$ over a defining width $\delta\psi$. This results in a total current per unit length along $\nabla\alpha$, no longer a surface current,

\begin{equation}
\mu_{0}\vec{k}(\alpha)\equiv \mu_{0}\int_{-\delta\psi/2}^{\delta\psi /2} \frac{\vec{j}}{\left| \nabla \psi \right| }d\psi = \mu_{0}j_{mn}e^{-im\alpha} \frac{\vec{B}}{\left| \nabla \psi \right| }.
\end{equation}

By Ampère's law, 
\begin{align}
\begin{split}
\mu_{0}\int_{-\delta\psi/2}^{\delta\psi /2} \frac{1}{\left|\nabla \psi\right|}\vec{j}_{\parallel}& \,d\psi= \int_{-\delta\psi/2}^{\delta\psi /2} \frac{1}{\left|\nabla \psi\right|} \nabla\times \vec{b} \,d\psi \\ 
 &= \int_{-\delta\psi/2}^{\delta\psi /2}  \frac{ \partial \vec{b}_{\alpha} }{ \partial \psi }  \,d\psi = \left[ \vec{b}_{\alpha} \right] _{-\delta\psi/2}^{\delta\psi /2}
\end{split}
\end{align}
From here we can continue to follow Boozer's logic, eventually relating $\left[ \vec{b}_{\alpha} \right] _{-\delta\psi/2}^{\delta\psi /2}$ to $\Delta_{mn}$ defined over the jump width $\delta\psi$, which we will write $\Delta_{mn}^{\delta \psi}$. This supports our assumption that the shielding current in response to a perturbation satisfies $\left| \vec{j} \right| \propto \left[ \vec{b}_{\alpha} \right] _{-\delta\psi/2}^{\delta\psi /2} \propto \Delta_{mn}^{\delta\psi}$. 

The careful reader will notice that this broadening of the definition of $\Delta_{mn}$ only holds in the immediate vicinity of  the rational surface, and is subject to the assumption that the perturbed current remains parallel to the equilibrium field. Thus, our approach to choosing a $\delta\psi$ will rely on scanning its magnitude and demonstrating both consistent placement relative to the variation in $\frac{ \partial b^{res}_{\psi} }{ \partial \psi }$ and a placement inside of any significant global kink structure in the perturbed equilibrium. 

\begin{figure}[ht]
\includegraphics[width=\linewidth]{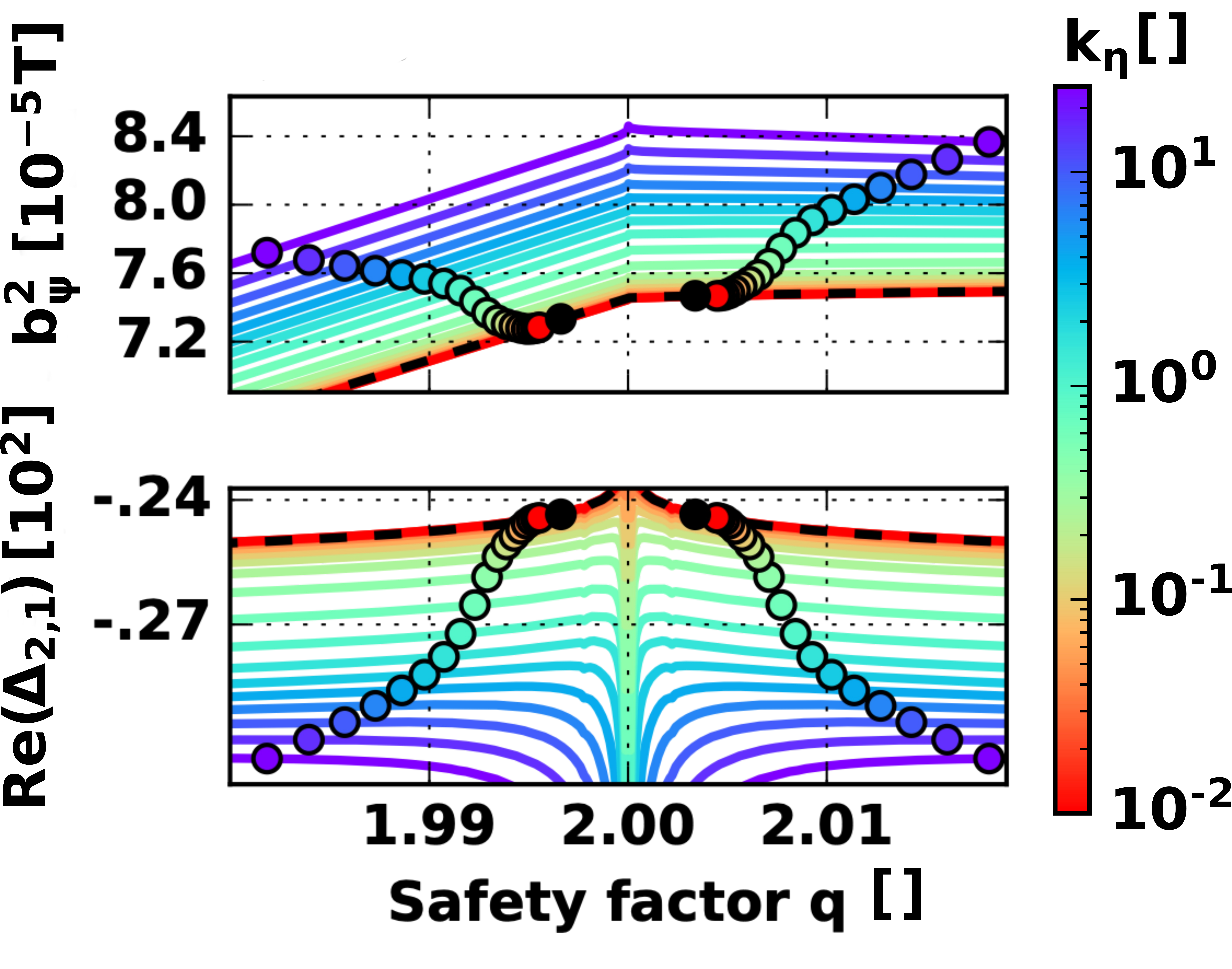}\protect\caption{Different discrete jump widths for different resistivities calculated using Eq.~\ref{eq:dpsi} are shown. The top panel shows these points on $b_\psi$ curves for $m=2$ in a scan of resistivity in a DIII-D-like scenario with $k_\eta=1$ representing the neoclassical resistivity. The dotted line is the same $b_\psi$ curve for $\eta=0$. The bottom panel shows for each resistivity value the $\Delta_{2,1}$ value that would be obtained by using a jump width at the given point, showing that until resistivity is relatively high, $\Delta_{2,1}$ remains relatively consistent in value. Note the small ($\sim 20\%$) range of the y-axis. \label{fig:deltamnchoice}}
\end{figure}

In GPEC, the half-width of the jump used for ideal cases defaults to $n\,\delta q = 5\times 10^{-4}$. When integrating with RDCON's Galerkin method, it is prudent to increase this, putting it outside the cells containing resonant basis terms to avoid the discontinuities in the derivative at the boundaries between resonant and non-resonant cells. The discontinuity in the radial derivative of $\vec{b}^{res}_{\psi}$ within this region can cause a $\Delta_{mn}^{\delta\psi}$ calculation to be numerically poorly-behaved. The present GGJ inner layer module also assumes that the inner layer width, and thus the width of the current sheet with it, grows inversely with Lundquist number, as $S^{-1/3}$ \cite{glasserComputationResistiveInstabilities2016a}. This is an inherent feature of the GGJ inner layer model in the limit of zero growth rate\cite{glasserResistiveInstabilitiesGeneral1975,glasserResistiveInstabilitiesTokamak1976}. Thus, we postulate that $\delta\psi$ should naturally vary with $S^{-1/3}$ as well. For this work, we have chosen, 

\begin{equation}
\delta\psi_{N} = 3.4\times 10^{-4} + \frac{S^{-1/3}}{2} \label{eq:dpsi}
\end{equation}

\noindent although, since this is a normalized poloidal flux width, the optimal values for both the starting value of $\delta\psi_{N}$ and the scale factor on its Lundquist number dependence will vary between dramatically differently sized devices. A scan in an ITER equilibrium that motivates this choice of $\delta\psi_{N}$ is seen in Fig.~\ref{fig:deltamnchoice}. It is clear here that the placement of the measuring points is both consistently outside the region in which $\Delta_{mn}^{\delta\psi}$ varies significantly due to the missing shielding current and placed as far inward as possible so as not to be disturbed by the large-scale mode structure of $b^{res}_{\psi}(\psi)$. This will allow us to calculate a $\Delta_{mn}$ for resistive perturbed equilibria, and therefore compare dominant mode spectra calculated from shielded flux to those calculated from resonant penetrated flux. This ability to calculate shielded flux dominant modes in resistive cases will also let us examine whether any differences between ideal and resistive coupling are due to inherent differences between $b_{sh}^{res}$ and $b_{pen}^{res}$ as resonant metrics or because of the resistive physics itself.

\section{\label{sec:behavior} Resonant metric behavior in resistive models}

\begin{figure}
    \includegraphics[width=\linewidth]{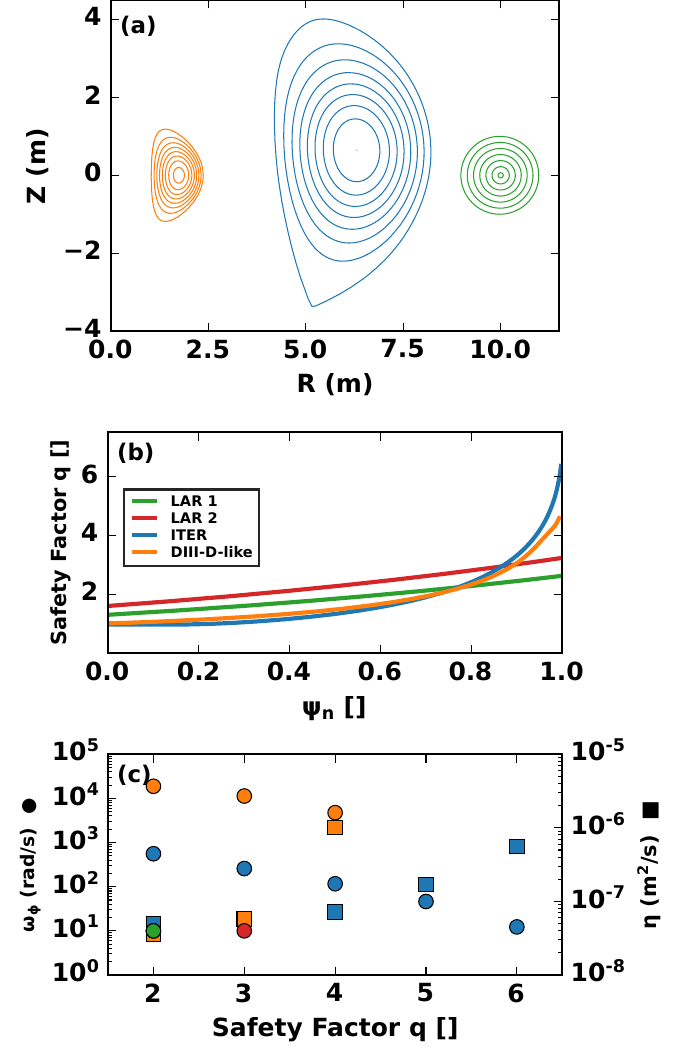}\protect\caption{Four test equilibria used in this paper. (a) shows cross-sections for each equilibrium, with the two-surface LAR not shown overlapping with the one-surface LAR. (b) contains the safety factor profiles for each equilibrium. (c) shows rotation (circles) and resistivity (squares) profiles for each equilibrium. \label{fig:equilibria}}
\end{figure}

\begin{table*}[]
\begin{tabular}{lllllllll}
\hline
Label & Device      & $B_T$  & $R$    & $a$    & $q_{95}$ & $\Omega_0$ & $\eta$ Source      & Equilibrium Source    \\ \hline
(a)   & LAR         & 1 T    & 10 m   & 1 m    & 2.53     & 10 rad/s   & Assumed            & Analytic  \\
(b)   & LAR         & 1 T    & 10 m   & 1 m    & 3.12     & 10 rad/s   & Assumed            & Analytic  \\
(c)   & ITER        & 2.65 T & 6.3 m  & 2 m    & 5.54     & 2 krad/s   & Jardin $\eta_{NC}$ & DINA \\
(d)   & DIII-D-like & 1.73 T & 1.74 m & 0.67 m & 3.75     & 50 krad/s  & Jardin $\eta_{NC}$ & Tokamaker \\ \hline
\end{tabular} \caption{Summary of equilibria used for metric and dominant mode comparisons. For the LAR cases (a) and (b), $\eta$ and $\Omega$ were the same on all rational surfaces, while the two gEQDSK equilibria (c) and (d) used different $\eta$ values based on a neoclassical equilibrium calculation and different $\Omega$ values based on a rotation profile. \label{tab:equilibria}}
\end{table*}

\begin{figure*}[ht]
    \includegraphics[width=\textwidth]{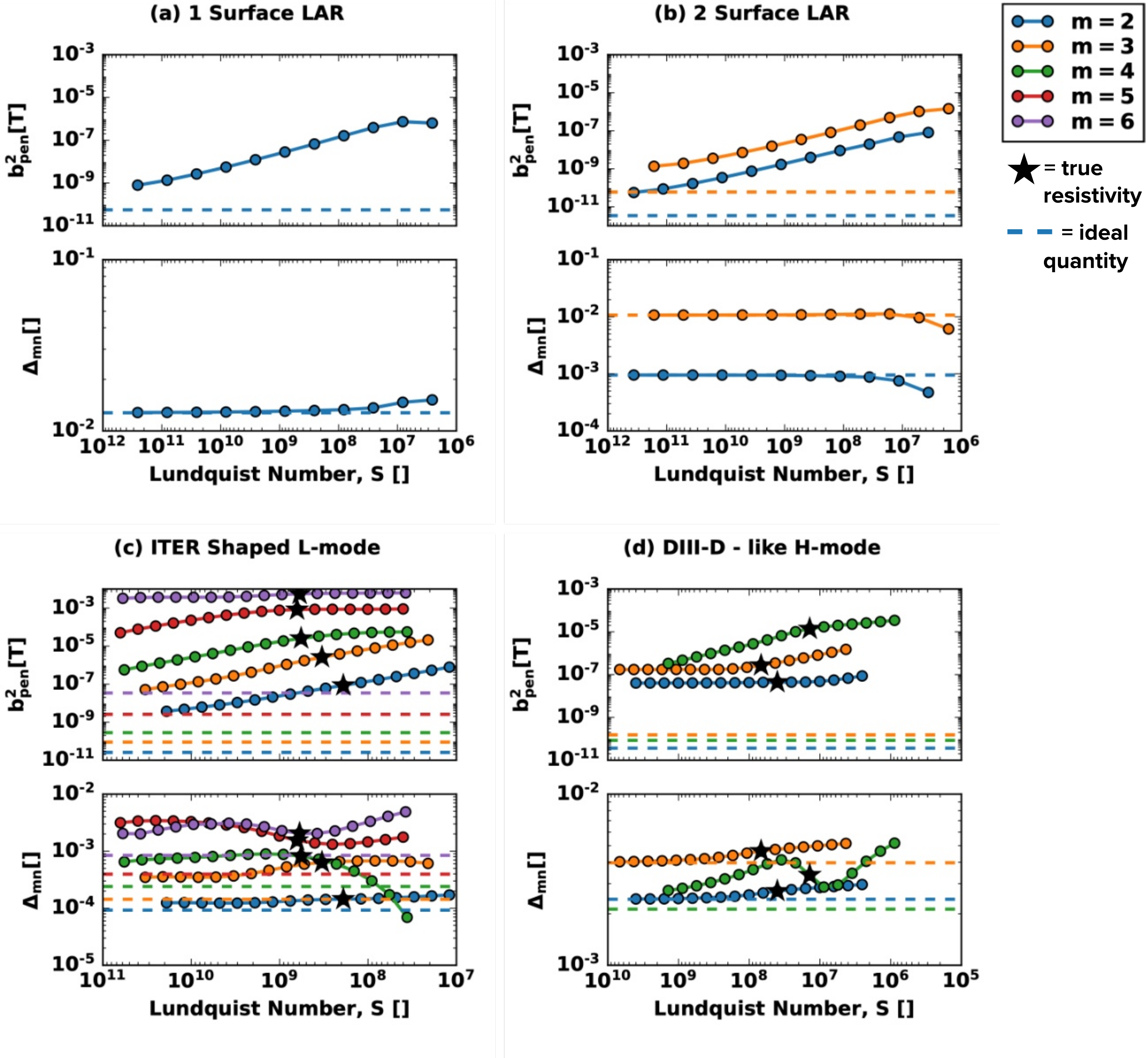}\protect\caption{Resonant penetrated field $b_{pen}^{res}$ and shielding current proxy $\Delta_{mn}$ at rational surfaces in the computational domain of different test equilibria. The consistent positive logarithmic slope in decreasing $S$ reveals the $b_{pen}^{res} \sim S^{-2/3}$ relationship. The value of $\Delta_{mn}$ in an ideal simulation of the same equilibrium is shown as a dotted line. For $b_{pen}^{res}$, the dotted line represents the numerical floor for the calculation; a penetrated field close to or lower than this value is impossible to calculate due to interpolation error. Details about the equilibria are found in Table~\ref{tab:equilibria} and Fig.~\ref{fig:equilibria}.\label{fig:metrics}}
\end{figure*}

This section examines how the two resonant metrics, the penetrated resonant field $b_{pen}^{res}$ and the resonant field derivative discontinuity $\Delta_{mn}$, vary with resistivity in several tokamak equilibria. We consider four cases: a Large Aspect Ratio (LAR) equilibrium with one rational surface in Fig.~\ref{fig:metrics} (a), a LAR equilibrium with two rational surfaces in Fig.~\ref{fig:metrics} (b), an ITER 5 MA L-mode equilibrium with low rotation in Fig.~\ref{fig:metrics} (c), and a DIII-D-like H-mode equilibrium with high rotation in Fig.~\ref{fig:metrics} (d). The LAR equilibria are analytical; the ITER equilibrium is a standard 5 MA half-field L-mode from DINA, and the DIII-D-like equilibrium was made with TokaMaker \cite{hansenTokaMakerOpensourceTimedependent2024,hansenOpenFUSIONToolkitOpenFUSIONToolkitV100beta52025} and is also used in \cite{burgessTearingStabilityPrediction2026}. We aim to understand the behavior of these two scalar metrics, with the dominant modes being examined in the following section. Further details of the equilibria used can be seen in Table~\ref{tab:equilibria} and Fig.~\ref{fig:equilibria}. 

The value of $b_{pen}^{res}$ increases monotonically with resistivity (and therefore decreasing Lundquist number $S$) until possible saturation in all four cases. This is consistent with the expectation that higher resistivity diffuses the shielding current sheet more, reducing its efficacy at suppressing the resonant penetrated field. In the two LAR equilibria, this increase is smooth and monotonic over the full range of S, just showing the beginning of saturation at the lowest S. In the ITER equilibrium, one observes an ordering to the saturation resistivity, with outer surfaces saturating at lower resistivities than inner surfaces. The fact that outer surfaces have lower rotation may contribute to this. This ordering presents in a different form in the DIII-D-like equilibrium, the penetrated field on all surfaces is not distinguishable from the numerical floor until we approach $S \approx 10^8$ for $m=2$ and $S \approx 10^7$ for $m=3,4$. This may be due to the much higher rotation assumed for this equilibrium (a two-segment linear profile), which rotates the forming island away from the external perturbation more quickly. In all cases, outer surfaces exhibit systematically higher values of $b_{pen}^{res}$, which is expected due to increasing distance from the dipole source in the EFC coils applying the field. The black star markers in Fig.~\ref{fig:metrics} (c) and (d) indicate the neoclassical resistivity values calculated using Eq.~17 of Jardin et al  \cite{jardinTSCSimulationOhmic1993} at the rational surfaces. 

In contrast to the penetrated field, $\Delta_{mn}$ displays much weaker dependence on resistivity. In the two LAR equilibria, $\Delta_{mn}$ remains visually indistinguishable from its ideal value except for at the lowest $S$. This is consistent with the expectation that adding a small resistivity to an ideal plasma equilibrium should diffuse and thicken the current sheet slightly, but not significantly change the total integrated shielding current, which is driven by an attempted change in flux. At higher resistivities, $\Delta_{mn}$ begins to deviate from its ideal value, as the resistivity becomes large enough to significantly alter the global structure of the plasma response. In this very high resistivity, low $S$ regime, our assumption that the inner layer is small compared to the space between layers can break down, invalidating the asymptotic matching approach. In the ITER and DIII-D-like equilibria, $\Delta_{mn}$ shows more complex behavior as resistivity is varied, reflecting the additional physics involved from additional rational surfaces and more poloidal mode coupling due to a lower aspect ratio and strong shaping. In these cases, $\Delta_{mn}$ may increase or decrease relative to its ideal value, and does not necessarily exhibit a clear monotonic trend or saturation like $b_{pen}^{res}$. Importantly, changes in the behavior of $\Delta_{mn}$ and $b_{pen}^{res}$ do not happen at the same resistivities. In the ITER case, $\Delta_{mn}$ does not appear to asymptote to the ideal value at high $S$ within the range displayed. However, at sufficiently high $S$, convergence toward the ideal value is observed, though this regime is not shown here due to numerical limitations.

Taken together, these results show that both resonant metrics behave in a well-defined but independent manner as resistivity is varied. Thus, scalar values of these metrics are not directly comparable measures of resonant drive in resistive perturbed equilibria. However, we will show in the following section that when used to calculate dominant mode spectra, both metrics yield closely similar results when evaluated within the same resistive model.

\section{\label{sec:efc} Resistivity's effect on error field correction}

\subsection{\label{sec:dommodes} Dominant mode spectra}

A simple way to approach assessing the coupling of resonant modes to external fields is to define coupling matrices that relate external fields or fluxes to a resonant metric. This approach has been used since the advent of the GPEC framework \cite{parkComputationThreedimensionalTokamak2007}, seen as $\overleftrightarrow{\mathcal{D}}$\cite{parkErrorFieldCorrection2008} in some early work and $\overleftrightarrow{C}$ most frequently\cite{parkErrorFieldCorrection2008,parkIdealPerturbedEquilibria2009,parkIdealPerturbedEquilibria2010,parkMDC19ReportAssessment2017}. This matrix is an inherent property of an equilibrium and can be used to evaluate the drive of different 3D perturbations. This matrix is frequently analyzed using singular value decomposition (SVD), and the singular vector with the highest singular value labeled the "dominant mode". Considering only the dominant mode is a fair approximation so long as there is strong separation in the magnitudes of the singular values. In this work, we define the coupling matrices using matrix vectors of half-area weighted (also known as energy-normalized) \cite{loganIdentificationMultimodalPlasma2016,loganEmpiricalScaling22020,loganRobustnessTokamakError2020,paz-soldanImportanceMatchedPoloidal2014,paz-soldanObservationMultimodePlasma2015,loganPhysicsBasisDesign2021,parkErrorFieldCorrection2008} surface-normal flux on the plasma edge $\tilde {\bf\Phi}^{ext}$ measured in Tesla, and resonant metrics ${\bf b}^{res}$ also in Tesla on each rational surface, such that
\begin{subequations}
    \begin{align}
    {\bf b}^{res}_{pen} &= {C_{\psi}}_{pen} \tilde {\bf \Phi}^{ext}\\
    {\bf b}^{res}_{sh} &= {C_{\psi}}_{sh} \tilde {\bf \Phi}^{ext}
    \end{align}
\end{subequations}
\noindent which have first singular vectors, i.e. dominant modes, ${\bf c}_{\psi_{pen}}^{(1)}$ and ${\bf c}_{\psi_{sh}}^{(1)}$, respectively. Both quantities being measured in Tesla makes the resulting coupling matrix unitless. Additionally, different works frequently use "core-dominant" \cite{pharrErrorFieldPredictability2024,loganRobustnessTokamakError2020,loganEmpiricalScaling22020,amoskovAssessment1Overlap2019} and "edge-dominant" \cite{loganMetricsExtrapolationResonant2025,luniaPredictedThresholdsRMP2025} modes to refer to singular vectors calculated by excluding surfaces outside or inside a certain $\psi_N$ range. This is motivated by the desire to capture perturbations that affect core confinement separately from those that affect edge stability; error field studies are concerned with core confinement, while RMP ELM suppression studies are concerned with edge stability. In this work, we only consider core-dominant modes, calculated by excluding surfaces outside $\psi_N = 0.9$.

\begin{figure}[ht]
    \includegraphics[width=\linewidth]{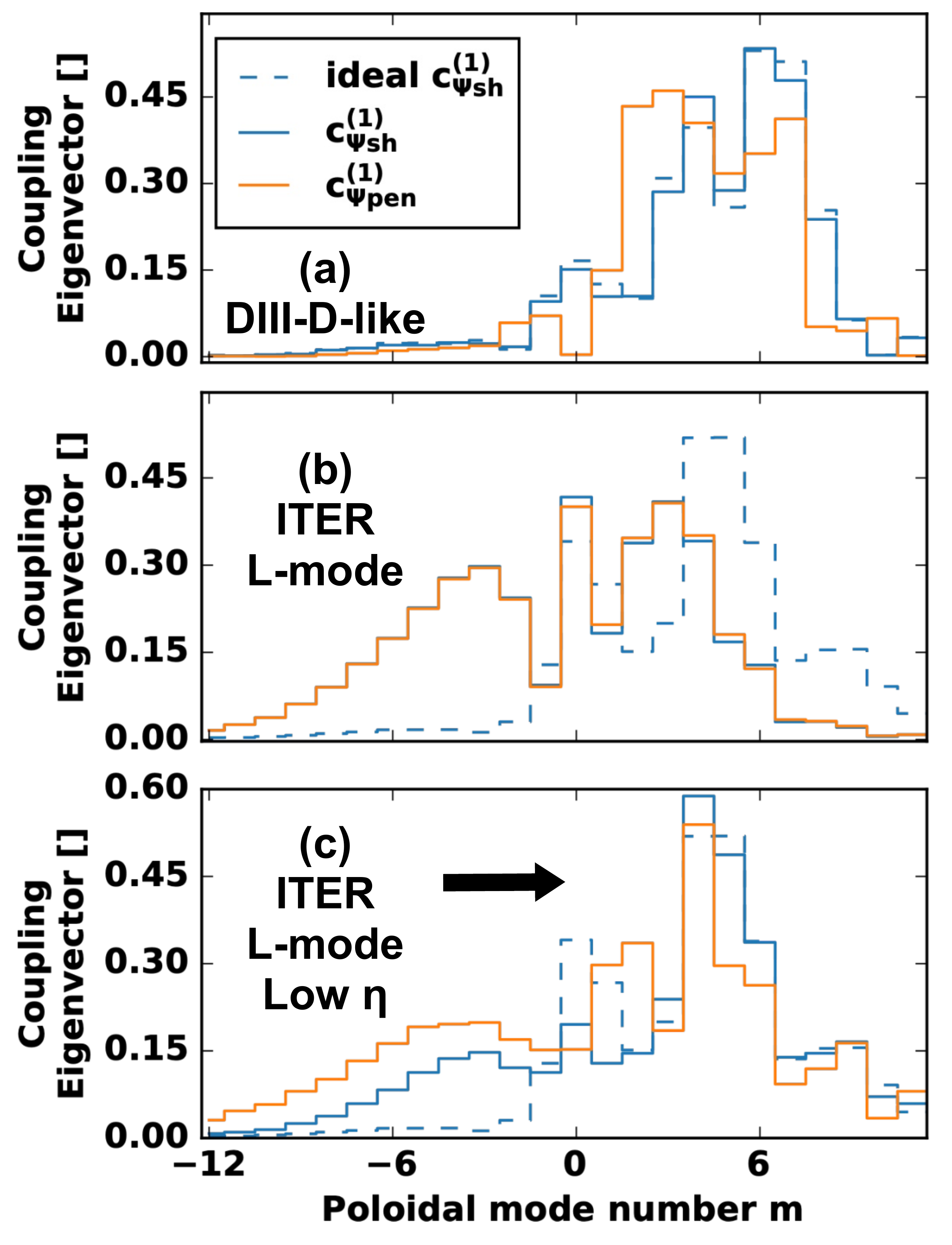}\protect\caption{Ideal (dotted) and resistive (solid) dominant modes from the (a) DIII-D-like and (b) ITER cases decomposed in Hamada coordinate poloidal modes. When resistivity is lowered to $10^{-3} \,\eta_{NC}$ (c), the resistive dominant modes shift rightward towards the ideal dominant mode. \label{fig:dommodes}}
\end{figure}

Fig.~\ref{fig:dommodes} (a) shows dominant modes for the DIII-D-like case, including both the shielded field dominant mode ${\bf c}_{\psi_{sh}}^{(1)}$ and the penetrated field dominant mode ${\bf c}_{\psi_{pen}}^{(1)}$ using the neoclassical resistivity values indicated by the black stars in Fig.~\ref{fig:metrics} (d). Also displayed, but mostly occluded is the ideal dominant mode using the shielded field metric. While not identical, the two resistive dominant modes are very similar in structure. The differences between the two resistive dominant modes may be attributable to the difficulty in accurately calculating $b_{pen}^{res}$ at this point in parameter space, as one can see from the black stars in Fig.~\ref{fig:metrics} (d) that resistivity is not high enough to have left the range where the $m=3$ field is flat. The numerical floor for $b_{pen}^{res}$ differs for ideal and resistive calculations due to differences in grid packing, potentially leading one to believe that this calculation is numerically limited. The accuracy of the formation of the pre-SVD coupling matrix relies on accurate ratios between metric values at different surfaces, supporting this inference. The resistive ${\bf c}_{\psi_{sh}}^{(1)}$ almost perfectly overlaps its ideal counterpart, as well. This indicates both that in this case, resistive physics does not significantly alter the dominant mode structure relative to ideal MHD, and that both resonant metrics yield similar dominant modes when evaluated within the same resistive model.

In contrast, Fig.~\ref{fig:dommodes} (b) shows dominant modes for the ITER case, again including both resistive dominant modes and the ideal shielded field dominant mode. The two resistive dominant modes overlap almost perfectly, both with considerable negative $m$ sensitivity and peaks at $m=0,3$; this confirms that despite the non-trivial differences in the variation of $\Delta_{mn}$ and $b_{pen}^{res}$, dominant modes calculated from them robustly represent the same physics. However, both resistive dominant modes show more lower $m$ coupling than the ideal dominant mode, which is peaked at $m=4-6$. This indicates that in this case, the inclusion of resistive physics modifies the dominant mode structure relative to ideal MHD, though both resonant metrics remain mutually consistent in the resistive model. This shows that resistive physics can modify the dominant mode spectrum relative to ideal calculations in certain regimes, such as this low-rotation ITER equilibrium. 

\begin{figure}[ht]
    \includegraphics[width=\linewidth]{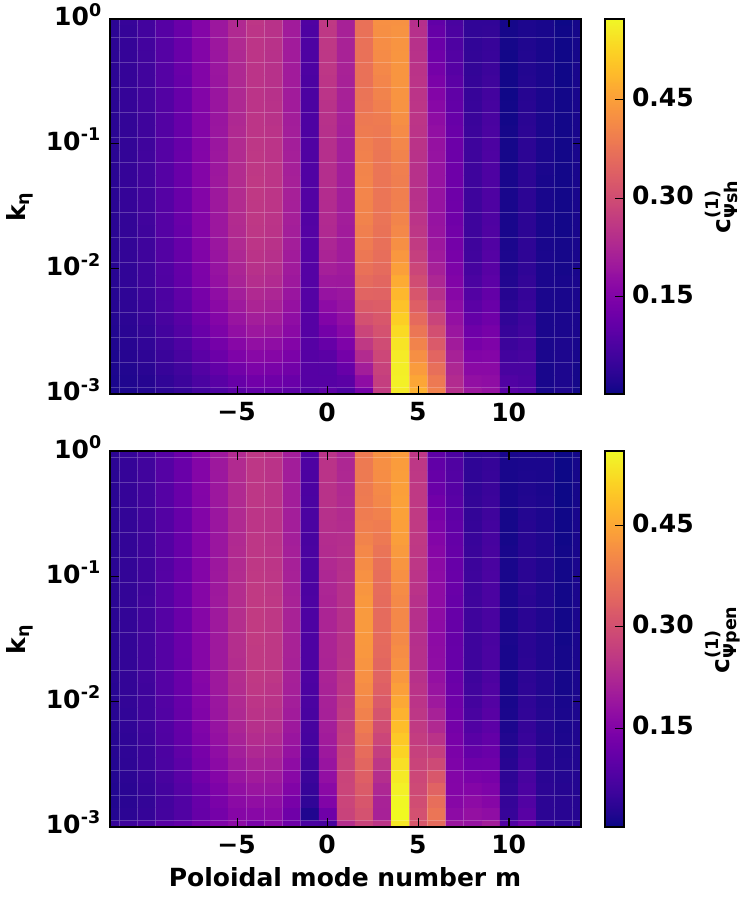}\protect\caption{Decreasing resistivity in the ITER simulation reveals that the peaks of both dominant modes shift toward higher $m$, converging towards the ideal ${\bf c}_{\psi_{sh}}^{(1)}$.\label{fig:dommodes_resscan}}
\end{figure}

Fig.~\ref{fig:dommodes} (c) shows the same ITER case but with resistivity on all surfaces reduced by a scale factor of $k_\eta = 10^{-3}$. The two resistive dominant modes can be seen to shift rightward, converging towards the ideal dominant mode peak. This shows that as resistivity is reduced, the dominant mode structure converges towards the ideal result, as expected. To demonstrate the robustness of this result, Fig.~\ref{fig:dommodes_resscan} shows a resistivity scan of the two dominant modes for the ITER case. As resistivity is reduced, both resistive dominant modes shift rightward smoothly.

This section yields two main results. First, both resonant metrics yield closely similar dominant mode spectra when evaluated within the same model. This means that decisions made for EFC coil design as well as the optimal distribution of currents between multiple EFC coils will be the same between calculations using either metric. This lends further confidence to model-based EFC and RMP ELM suppression designs. This is a novel result and also validates the use of $b_{pen}^{res}$ as a resonant metric to compute coupling matrices and dominant modes in other settings, such as M3D-C1 and MARS-F calculations. Second, the inclusion of resistive physics can modify the dominant mode coupling spectrum relative to ideal calculations. This is particularly notable for ITER-relevant L-mode equilibria with low rotation, which are particularly sensitive to error fields. Furthermore, resistivity lowering the $m$ of the peak coupling may increase error field risk as tilt and shift coil misalignments tend to produce lower $m$ perturbations. 

\subsection{\label{sec:ccs} Optimal RMP coil phasings}

Ideal GPEC has been shown in the past to accurately predict optimal RMP current phasings for EFC \cite{paz-soldanSpectralBasisOptimal2014,paz-soldanImportanceMatchedPoloidal2014} and RMP ELM suppression \cite{huNonlinearTwofluidModeling2021, parkGeneralizedPerturbedEquilibrium2018, paz-soldanObservationMultimodePlasma2015,yangTailoringTokamakError2024}. This is expected for equilibria with low resistivity and high rotation, where the ideal dominant mode is preserved, like in the DIII-D-like case. However, the different dominant mode in ITER seen in Fig.~\ref{fig:dommodes} (b) indicates that the ideal dominant mode may not be a good approximation for the resistive dominant mode in some cases. Thus, an experiment on a device could be conducted to validate the resistivity-dependence of the dominant mode spectrum, and to determine whether the ideal dominant mode is a good approximation for the resistive dominant mode in ITER-relevant L-mode equilibria. This would be done by finding an equilibrium such as the ITER case, and then finding the relative phasings of multiple sets of EFC coils that cause mode locking at the lowest current amplitude. This would be compared to the ideal and resistive dominant mode predictions, and the resistivity-dependence of the dominant mode spectrum could be validated with other models such as MARS-F and linear M3D-C1.

\begin{figure}[ht]
    \includegraphics[width=\linewidth]{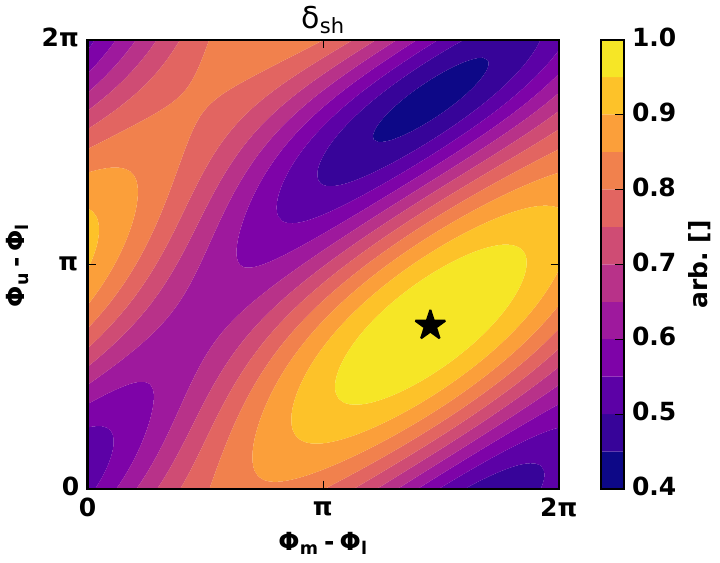}\protect\caption{The optimal relative phasing between ITER's three sets of EFC coils using the ideal MHD model and shielding metric. The phasings that maximize the overlap of the applied field with the dominant mode, $\delta_{sh}$, is shown with a black star. \label{fig:idealphasing}}
\end{figure}

\begin{figure}[ht]
    \includegraphics[width=\linewidth]{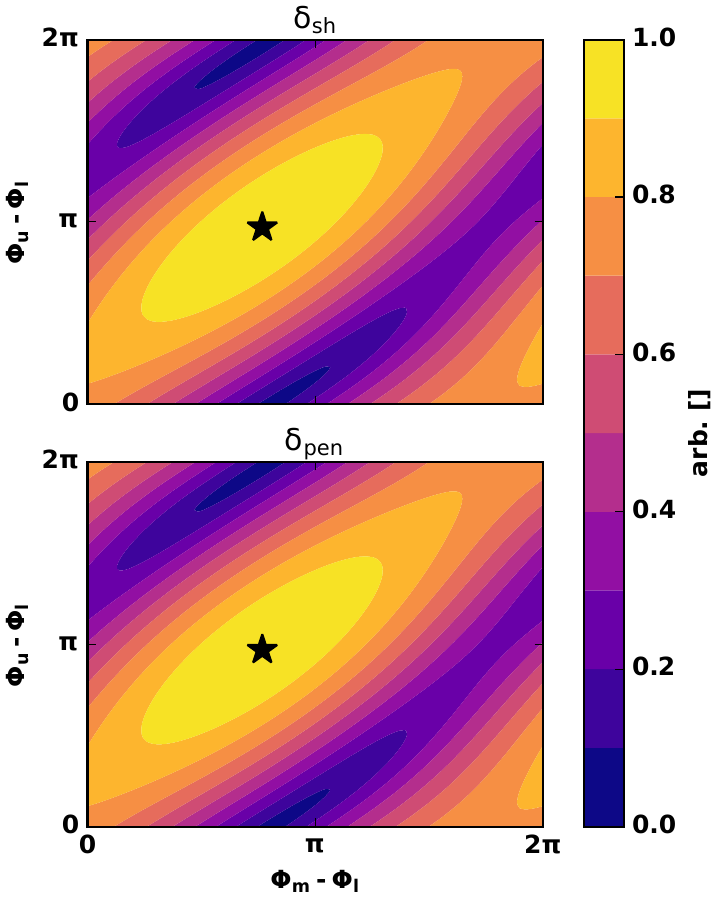}\protect\caption{The overlap with shielding ($\delta_{sh}$) and penetrated field ($\delta_{pen}$) dominant modes are shown as calculated from perturbed equilibria with resistive layers. The optimal relative phasings are indicated with black stars. The optimal relative phasings are the same using either metric, but the resistive model predicts a significantly different set of phases than the ideal model.\label{fig:resphasing}}
    
\end{figure}

These predictions are made by calculating the maximum overlap of the applied field from each coil with the dominant mode across relative phasings, given by the inner products, 
\begin{align*}
\delta_{sh} &= \frac{1}{B_T}{\bf c}_{\psi_{sh}}^{(1)} \cdot \tilde {\bf \Phi}^{ext} \\ 
\delta_{pen} &= \frac{1}{B_T}{\bf c}_{\psi_{pen}}^{(1)} \cdot \tilde {\bf \Phi}^{ext}. 
\end{align*}
$\delta_{sh}$ is shown calculated using the ideal dominant mode in Fig.~\ref{fig:idealphasing}, and $\delta_{sh}$ and $\delta_{pen}$ are shown calculated from the resistive dominant mode in Fig.~\ref{fig:resphasing}.  While the two resistive metrics yield the same optimal relative phasing, the ideal and resistive dominant modes yield different optimal relative phasings. The relative upper-lower phasings based on $\delta_{sh}$ differ by roughly 44 degrees, while the middle-lower phasings differ by a larger 124 degrees. Applying the currents with optimal phasings from the ideal dominant mode would result both in weaker (68\% strength) coupling to the resistive dominant mode and a plasma response phase mismatch of 94 degrees --- a significant difference that would render attempted EFC using model-derived coil phases ineffective.

This difference is due to fundamentally different coupling between the field applied by ITER's EFC coils and the dominant mode in the ideal and resistive models. This is an experimentally testable prediction, and could be validated on ITER or other devices with multiple sets of EFC coils, in a low-rotation equilibrium with strong resistivity.

\section{\label{sec:summary} Summary}

We present a study of two metrics for resonant drive, the resonant field derivative discontinuity $\Delta_{mn}$ and the resonant penetrated field $b_{pen}^{res}$, in resistive perturbed tokamak equilibria using the resistive GPEC model with a GGJ inner layer. We show in Figs.~\ref{fig:deltamnchoice} and \ref{fig:metrics} that both metrics remain well-defined for thin inner layers using a physically motivated finite jump width for calculating $\Delta_{mn}$. We examine how these metrics behave as resistivity is varied in several equilibria, showing that $b_{pen}^{res}$ increases monotonically with resistivity until possible saturation, while $\Delta_{mn}$ shows more complex behavior. 

This work serves as a strong argument that $b_{pen}^{res}$ is a robust metric for resonant drive in resistive perturbed equilibria, yielding results consistent with the more traditional $\Delta_{mn}$ metric. Future work should examine how these results generalize to other inner layer models, and nonlinear simulations could be used to validate the predictions of linear resistive models.

Both metrics can be used to calculate dominant mode spectra by taking the SVD of coupling matrices relating external perturbations to the resonant metrics. These dominant mode spectra are found to be very similar as seen in Fig.~\ref{fig:dommodes}, indicating that both metrics quantify the same underlying physics of resonant drive. This supports the validity of dominant modes calculated using resonant penetrated field calculated by other resistive codes. However, we see that the inclusion of resistive physics can modify the dominant mode spectrum relative to an ideal calculation in some regimes, in this case an ITER low-rotation L-mode equilibrium. The resistivity-dependence of the dominant mode should be validated with other models such as MARS-F and linear M3D-C1, and can be tested via experimental validation of altered optimal EFC coil phasings. Furthermore, especially given the lower $m$ peaking of the resistive dominant mode discussed in Sec.~\ref{sec:dommodes}, a study should be conducted on core-dominant modes in ITER and fusion pilot plants to evaluate whether the resistivity dependence has implications for construction tolerancing.

\section*{Acknowledgments}
This work was supported by US DOE awards DE-SC0022272 and DE-SC0024898. This work was also supported by the National Research Foundation of Korea (NRF), Ministry of Science and ICT, No. RS-2023-00281276 and by R\&D Program of “Optimal Basic Design of DEMO Fusion Reactor, CN2502-1” through the Korea Institute of Fusion Energy(KFE) funded by the Government funds.
Thanks to Zhirui Wang and Alan Glasser for significant help with RDCON and DeltaC/RMATCH, to Stuart Benjamin for important sensitivity analysis of RDCON's $\Delta'$ calculations, and to Priyansh Lunia for text review. 

\section*{References}

\bibliographystyle{iopart-num} 
\bibliography{bibtexrefs.bib}

\end{document}